\documentclass{article}
\usepackage{style/spconf,amsmath,graphicx}
\usepackage{xcolor}
\usepackage{booktabs}
\usepackage{multirow}
\usepackage{caption}
\usepackage{subcaption}
\usepackage{enumitem}
\usepackage[normalem]{ulem}
\usepackage{cite}
\usepackage{float}
\usepackage{amssymb}
\usepackage{pifont}
\newcommand{\cmark}{\ding{51}}%
\newcommand{\xmark}{\ding{55}}%

\title{An ASR-free fluency scoring approach with Self-Supervised Learning}

\name{Wei Liu$^{1,*}$\thanks{$^{*}$ This work was done during an internship at ByteDance.}, Kaiqi Fu$^{2}$, Xiaohai Tian$^{2}$, Shuju Shi$^{2}$, Wei Li$^{2}$, Zejun Ma$^{2}$ and Tan Lee$^{1}$}
\address{
$^{1}$Department of Electronic Engineering, The Chinese University of Hong Kong \\
$^{2}$ByteDance}

\begin{document}
\ninept
\maketitle
\begin{abstract}
A typical fluency scoring system generally relies on an automatic speech recognition (ASR) system to obtain time stamps in input speech for the subsequent calculation of fluency-related features or directly modeling speech fluency with an end-to-end approach. This paper describes a novel ASR-free approach for automatic fluency assessment using self-supervised learning (SSL). Specifically, wav2vec2.0 is used to extract frame-level speech features, followed by K-means clustering to assign a pseudo label (cluster index) to each frame. A BLSTM-based model is trained to predict an utterance-level fluency score from frame-level SSL features and the corresponding cluster indexes. Neither speech transcription nor time stamp information is required in the proposed system. It is ASR-free and can potentially avoid the ASR errors effect in practice. Experimental results carried out on non-native English databases show that the proposed approach significantly improves the performance in the ``open response" scenario as compared to previous methods and matches the recently reported performance in the ``read aloud" scenario.


\end{abstract}

\section{Introduction}
\label{sec:intro}
Speech fluency/disfluency is an important indicator of language proficiency level in second language (L2) learning~\cite{housen2009complexity}. It is usually described and measured in terms of a set of time-domain measures, for example, speech rate,  silence duration, filled pauses, and self-correction, etc~\cite{kormos2004exploring, cucchiarini2002quantitative, ellis1994study, meng2009developing}. Automatic scoring of fluency, as such, serves as an essential module in computer-aided language learning (CALL) systems.

There have been numerous studies on automatic speech fluency assessment. They include the conventional hand-crafted features based approaches and the more recent end-to-end approaches.
In the features based approaches, various types of fluency-related features, e.g. statistics of speech breaks \cite{loukina2019automated, deng2020automatic, fontan2020using}, speech rate \cite{loukina2019automated, deng2020automatic, fontan2020using, van2015automatically, kjellgren2016convolutional}, filled pauses \cite{deng2020automatic} and goodness of pronunciation (GOP) \cite{mao2019nn, zhang2021multilingual}, are computed from time stamps provided for input speech, e.g.,  beginning and end time of words, phonemes. Different machine learning models, e.g., support vector machine (SVM) \cite{mahesha2013approach, deng2020automatic} and feed-forward neural networks \cite{fontan2020using}, were trained to predict fluency scores from the extracted features. End-to-end approaches were developed to predict the utterance-level fluency scores directly from phone-level speech features. 
In~\cite{gong2022transformer}, a multi-aspect and multi-granularity Transformer was adopted to capture the dynamic changes of phone-level pronunciation-related features. A follow-up work attempted to further improve the system by using multi-view representations that comprise additional prosodic and SSL speech features~\cite{chao20223m}. 
In~\cite{fu22b_interspeech}, Bi-directional Long Short Term Memory (BLSTM) is adopted to predict fluency scores from a sequence of phone-level deep features (usually extracted from the bottleneck layer of a DNN acoustic model) and its corresponding phone duration features.

The above approaches share a common requirement that an ASR system is needed for obtaining the time stamps of the speech segments concerned. The forced-alignment approach is adequate for the ``read aloud" scenario where text prompts are given. In the ``open response" scenario,  the responses are spontaneous speech for which the content is not given. An ASR system is needed in this case.  Errors made in the ASR process would inevitably affect the performance of fluency scoring. This motivates the investigation of ASR-free approaches \cite{chung2017spoken, fontan2020using}. In~\cite{chung2017spoken}, Convolution Neural Network (CNN) was applied to learn a direct mapping from raw waveform to fluency score.  An unsupervised Forward-Backward Divergence Segmentation (FBDS) algorithm \cite{andre1988new} was proposed to obtain time boundaries of pseudo-syllables and silent breaks,  and hand-crafted fluency features were derived for fluency scoring in \cite{fontan2020using}. Both studies showed promising results and suggested the feasibility of fluency scoring in an ASR-free manner. However, the datasets used were relatively small (less than 500 utterances). The robustness of the framework needs to be further examined with larger datasets.

Recently, self-supervised learning (SSL) based speech models, e.g. wav2vec2.0~\cite{baevski2020wav2vec}, HuBERT~\cite{hsu2021hubert}, have been shown capable of learning meaningful representations for various downstream tasks~\cite{mohamed2022self}.
These models also demonstrated effectiveness in unsupervised acoustic unit discovery, where acoustic units obtained by performing K-means clustering of SSL features are aligned well with phonemic units identified by human experts \cite{baevski2021unsupervised}. 
This paper proposes a novel approach to ASR-free fluency scoring based on unsupervised speech segmentation with SSL representations. A speech signal is encoded into frame-level SSL representations using wav2vec2.0. K-means clustering is applied to the SSL representations and a cluster index is assigned to each frame. A BLSTM-based scorer is trained to predict an utterance-level fluency score from frame-level features and cluster information. Neither transcriptions nor time stamp information is required in the system. It is ASR-free and can potentially improve performance by alleviating the impact of ASR errors.


\begin{figure*}[ht!]
	\centering
	\includegraphics[width=165mm]{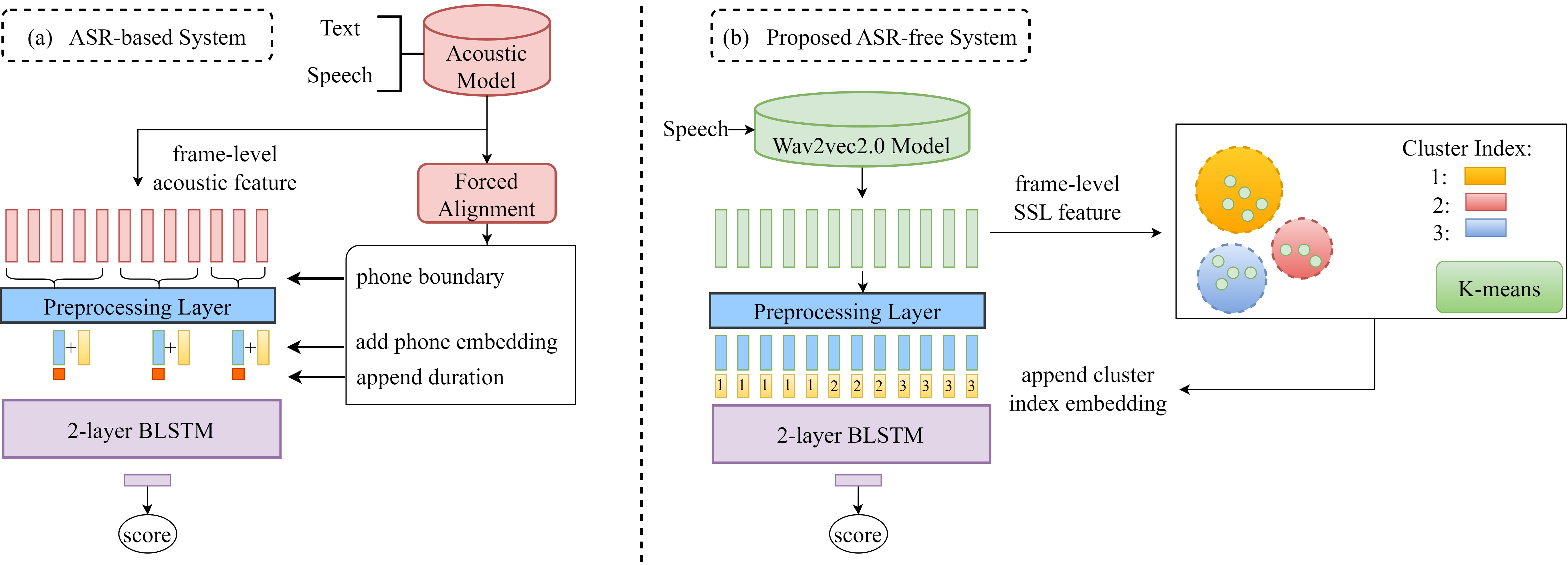}
	\caption{Block diagram of system architectures. (a) illustrates the ASR-based fluency scoring system, which uses an acoustic model to derive time stamps of speech units. In the ``open response" scenario, an additional ASR system is used to obtain the speech transcription. (b) illustrates the proposed ASR-free fluency scoring system, which does not require speech transcription and time stamp information. }
	\label{fig:system_overview}
\end{figure*}

\section{Method}
\label{sec:method}

\subsection{ASR-based Fluency Scoring System}
\label{sec:segment_level system}



In~\cite{fu22b_interspeech}, an ASR-based fluency scoring system was reported with encouraging results in the ``read aloud" scenario. To extend the system to the ``open response" scenario,
an additional step of speech-to-text conversion can be adopted. Then the remaining procedure would be the same as in the ``read aloud" scenario. This system is regarded as one of the baseline systems in the present study.

As shown in Figure~\ref{fig:system_overview} (a), ASR-based fluency scoring involves two stages: feature extraction and fluency scoring. Given a speech utterance, frame-level acoustic features denoted as $\mathbf{X_f} \in \textbf{R}^{\text{D1} \times \text{T}}$ are first extracted from the acoustic model. $\text{D1}$ and $\text{T}$ denote acoustic feature dimension and the number of frames, respectively. Forced alignment is performed on the input utterance to obtain time stamps at phone level, e.g., the start and end time of each phone and silence incidence. Subsequently, phone-level acoustic features and phone duration are extracted, which are denoted as $\mathbf{X_s} \in \textbf{R}^{\text{D1} \times \text{N}}$ and $\mathbf{t}\in \textbf{R}^{\text{1} \times \text{N}}$, respectively.  The corresponding phoneme sequence is denoted as $\mathbf{e}\in \textbf{R}^{\text{1} \times \text{N}}$, where $\text{N}$ is the number of phones.

At the fluency scoring stage, phone-level acoustic features are first transformed into a compact feature space by a preprocessing layer $\mathcal{P}$,
\begin{equation}
\label{eq:preprocessing}
\mathbf{H_s} = \mathcal{P}(\mathbf{X_s}),
\end{equation}
where $\mathbf{H_s} \in \textbf{R}^{\mathbf{D_h} \times \mathbf{N}}$. $\mathbf{D_h}$ denotes the hidden representation dimension.

Then, we project the phoneme IDs, $\mathbf{e}$, to the phoneme embedding vectors $\mathbf{E}\in \textbf{R}^{\mathbf{D_h} \times \mathbf{N}}$. The sum of $\mathbf{E}$ and $\mathbf{H_s}$ is concatenated with corresponding phoneme duration features $\mathbf{t}$ as the input of BLSTM-based fluency scorer $\mathcal{S}$ for fluency assessment. 
\begin{equation}
\label{eq:scoring}
\mathbf{\hat{y}} = \mathcal{S}([ \mathbf{E}+\mathbf{H_s}; \mathbf{t} ]),
\end{equation}
where $\mathbf{\hat{y}}$ is the predicted fluency score. The detailed model structure can be found in~\cite{fu22b_interspeech}.


\subsection{The Proposed ASR-free Fluency Scoring System}
\label{subsec:proposed_system}

\subsubsection{Unsupervised Unit Discovery on Self-supervised Feature}
\label{subsec:motivation4ssl}
Self-supervised learning (SSL) has been applied to learn representations from large-scale unlabeled data~\cite{baevski2020wav2vec, hsu2021hubert}. It has shown great success in many downstream tasks, such as automatic speech recognition, speaker verification, unsupervised unit discovery, etc. 
In~\cite{baevski2020wav2vec}, it was found that the learned discrete latent speech representations are highly related to phonemic units. The work presented in~\cite{hsu2021hubert} also examined the correlation between learned units by K-means clustering and phonetic units obtained by forced-alignment. A recent work applied SSL features on unsupervised speech recognition \cite{baevski2021unsupervised}, with the aim of learning the mapping from SSL audio representations to phonemes without using labelled data. The study demonstrated that leveraging high-quality speech representations learned from the SSL model, e.g. wav2vec2.0, is able to segment speech into meaningful units with a simple clustering method. 

In this study, we propose an ASR-free fluency scoring framework based on the hypothesis that speech segments obtained by unsupervised unit discovery contain the information related to the phoneme boundaries, which is able to replace the time stamps of phonemes obtained by forced alignment.

\subsubsection{System Architecture}
As shown in Figure \ref{fig:system_overview} (b), given a speech utterance, wav2vec2.0 is applied to encode the time-domain signal into
frame-level representations, $\mathbf{X_f} \in \textbf{R}^{\text{D2} \times \text{T}}$, where $\text{D2}$ and $\text{T}$ are the dimensions of SSL features and the number of frames, respectively. 
Instead of explicitly obtaining phone-level information via forced alignment (as mentioned in Section~\ref{sec:segment_level system}), K-means algorithm is applied to the frame-level SSL features to assign each frame with a unique cluster index $\text{i} \in [1, K]$, where $K$ is the number of clusters. 

Different from the system mentioned in Section~\ref{sec:segment_level system}, where the operation is based on phone-level features, the frame-level feature $\mathbf{X_f}$ is used as the input of preprocessing layer. Hence, the Equation~\ref{eq:preprocessing} is rewritten as
\begin{equation}
\label{eq:preprocessing2}
\mathbf{H_f} = \mathcal{P}(\mathbf{X_f}),
\end{equation}
where $\mathbf{H_f} \in \textbf{R}^{\mathbf{D_h} \times \mathbf{T}}$. $\mathbf{D_h}$ and $\text{T}$ denote the dimensions of hidden representations and the number of frames, respectively.

Then, we project the cluster indexes, $\textbf{i}$, to the embedding vectors $\mathbf{I}\in \textbf{R}^{\mathbf{D_I} \times \mathbf{T}}$ and concatenate with the hidden representations $\mathbf{H_f}$ as the fluency scorer input for fluency score prediction, expressed as
\begin{equation}
\label{eq:scoring2}
\mathbf{\hat{y}} = \mathcal{S}([ \mathbf{H_f}; \text{I} ]).
\end{equation}

\section{Experimental Setup}
\label{sec:exp_setup}

\subsection{Dataset}

\begin{table}[t!]
\centering
\caption{The summary of statistics of corpus ByteQA.}
\begin{tabular}{c|ccc}
\toprule
Data sets          & train & dev   & test  \\ \midrule
\# of Spks         & 161   & 53    & 55    \\
\# of Utts         & 4404  & 1395  & 1451  \\ \midrule
Avg \# of words    & 32    & 33    & 34    \\
Avg duration (sec) & 19.8  & 21.0  & 20.9  \\
Max duration (sec) & 172.5 & 155.4 & 153.2 \\ \bottomrule
\end{tabular}
\label{tab:dataset}
\end{table}
The ByteQA speech database is used for fluency scoring assessment in this study. ByteQA is an internal dataset at ByteDance that contains 7250 English utterances by 269 adult Mandarin learners collected under the ``open response" scenario. The data collection process is as follows: Each learner starts with watching a short (0.5-2 minutes in duration) video in English of which the topic is non-academic and more about business scenarios, and the learner will then answer a few questions about the video in speech. Two linguistic experts rated each response at a scale of 0-4 with 0 being not fluent at all and 4 being very fluent. A third linguistic expert is included to re-assess the items with different scores between the previous two raters, and the scores given by the third rater are taken as final.  A detailed description of ByteQA is given in Table \ref{tab:dataset}.  


\subsection{ASR-related Setup}
\label{sec:asr_setup}
\begin{itemize}[leftmargin=*]
    \item \textbf{Acoustic Model:} The architecture in~\cite{zhang2018deep} is adopted which is a hybrid system of feedforward sequential memory networks (FSMN) and hidden Markov models (HMM), i.e., DFSMN-HMM. The deep FSMN architecture consists of 2 convolution layers and 24 FSMN layers followed by two fully connected (FC) layers. The input features are 39-dim Mel-frequency cepstral coefficients (MFCCs) and the acoustic model is trained on a corpus of about 5,000 hours of English speech, including an internal corpus of 4K hours of non-native speech by Mandarin learners and 960 hours of native English speech from the Librispeech corpus~\cite{panayotov2015librispeech}.
    \item \textbf{Speech Transcriptions:} Speech transcriptions of the ByteQA dataset are obtained separately in two ways, human annotations (as the ground truth) and ASR transcription. The ASR system is trained using 40k hours of English speech data (including both native and non-native speech) with the RNN-T architecture~\cite{graves2012sequence}, which yields $12.2\%$ word error rate (WER) on ByteQA.
\end{itemize}

\subsection{System Setup for Baseline Models and the Proposed Model}
\begin{itemize}[leftmargin=*]
    \item \textbf{Waveform + CNN} \cite{chung2017spoken}: Raw waveforms are directly used as input to a three-layer 1D-CNN model for fluency score prediction. The filter size and stride parameters of the three-stacked CNN layers are empirically decided in \cite{chung2017spoken} to capture the short-term, mid-term, and long-term characteristics of the speech signal.
    
    \item \textbf{FBDS} \cite{fontan2020using}: The FBDS algorithm is employed to segment speech recordings into sub-phonemic units, followed by automatic clustering of the sub-phonemic units into higher-level pseudo-syllables and silent breaks. Hand-crafted fluency-related features are then calculated and fed to a three-FC-layer NN for fluency prediction.
    \item \textbf{Handcrafted features} \cite{deng2020automatic}:  
    The model input comprises three prosodic features (i.e., speech rate, the ratio of number of pauses to number of syllables, and silence rate) and four disfluency-related features (e.g., filled pauses and word fragments related). Note that we use syllables instead of moras \cite{deng2020automatic}, given the target language is English. SVM is then applied to predict the fluency scores where grid-search is utilized for finding the optimal kernel function and hyperparameters over the validation set.
    \item \textbf{Deep feature+ASR-based} \cite{fu22b_interspeech}:  Figure \ref{fig:system_overview} (a) shows the architecture of this framework. The model inputs are 512-dim deep features extracted from the penultimate layer of the acoustic model (Section \ref{sec:asr_setup}). 
    
    \item \textbf{Proposed SSL+ASR-free}: The main network architecture basically follows the ASR-based fluency scoring system in \cite{fu22b_interspeech}, including a preprocessing layer and a two-layer BLSTM. The preprocessing layer consists of a linear layer followed by a \textit{LayerNorm} and Tanh activation function. All the intermediate hidden dimension is set as $32$. Following \cite{baevski2021unsupervised}, the $15$-th layer representation of wav2vec2.0 Large~\footnote{\scriptsize{https://dl.fbaipublicfiles.com/fairseq/wav2vec/wav2vec\_vox\_new.pt}} is selected as the SSL input features. The number of K-means clusters is set as $50$ and the respective cluster embedding dimension $\mathbf{D_I} = 6$, which are decided by preliminary experiments.  For network training, the maximum epoch is set as $50$ with early stopping criteria ($patience=7$). With learning rate $=0.002$ and batch size $=32$, the Adam \cite{kingma2014adam} is utilized to optimize the mean square error (MSE) loss between the normalized ground-truth scores and the predicted scores with a range of [-1,1] (using Tanh activation function as output). 
\end{itemize}


\begin{table}[t!]
\centering
\caption{Performance comparison between previous methods and the proposed one on the ByteQA dataset. Text source indicate what kind of transcription is used for forced alignment.}
\resizebox{\columnwidth}{!}{

\begin{tabular}{cc|cc|c}
\toprule
\multicolumn{2}{c|}{Methods}                                                                                                                         & ASR-free              & Text source           & PCC   \\ \midrule
\multicolumn{1}{c|}{\multirow{4}{*}{\textrm{I}}}   & Wavform+CNN     \cite{chung2017spoken}                                                            & \cmark & \xmark & 0.575 \\
\multicolumn{1}{c|}{}                                               & FBDS  \cite{fontan2020using}                                          & \cmark & \xmark & 0.549 \\
\multicolumn{1}{c|}{}                                               & 
Handcrafted features \cite{deng2020automatic}                                     & \xmark & human                 & 0.690 \\
\multicolumn{1}{c|}{}                                               & Deep feature + ASR-based \cite{fu22b_interspeech}                                               & \xmark & human                 & 0.759 \\ \midrule
\multicolumn{1}{c|}{\multirow{2}{*}{\textrm{II}}}  & SSL+ASR-based                                                                    & \xmark & human                 & 0.812 \\
\multicolumn{1}{c|}{}                                               & SSL+ASR-based                                                                    & \xmark & asr                   & 0.791 \\ \midrule
\multicolumn{1}{c|}{\textrm{III}} & \textbf{Proposed SSL+ASR-free}                                                                     & \cmark & \xmark & \textbf{0.797} \\ \bottomrule
\end{tabular}

}
\label{tab:compare_with_baselines}
\end{table}

\section{Results and Analysis}
\label{sec:exp}
The effectiveness of the proposed method is evaluated on both the ``open response" and the ``read aloud" scenarios. The experimental results are given in Table~\ref{tab:compare_with_baselines} and Table~\ref{tab:read_aloud}.
To assess the system performance, Pearson correlation coefficient (PCC) between machine-predicted scores and human scores was used in our experiments. A higher PCC value indicates better system performance.

\subsection{Experimental Results in the ``Open Response" Scenario}
\subsubsection{Comparative studies across competitive methods}
We first compare the performance of the proposed ASR-free approach with various baseline methods. As shown in Table~\ref{tab:compare_with_baselines} group (\textrm{I}), 
the proposed approach performs significantly better than the two ASR-free baselines with a PCC value of 0.797. This also implies that speech segmentation using unsupervised unit discovery in the current proposed approach could be superior to the FBDS and Waveform+CNN baselines.

We then compare our SSL+ASR-free approach with the baselines, which utilize an ASR module to perform forced alignment as presented in the group (\textrm{I}) of Table~\ref{tab:compare_with_baselines}.
The results suggest that the proposed method consistently outperforms the baseline models, which utilize an ASR module irrespective of the inputs of the ASR-based approaches (handcrafted features v.s. deep features). 

To further investigate the effect of omitting the ASR module in the current framework, we conducted experiments combining the SSL features and ASR-based system (Section \ref{sec:segment_level system}), and the results are given as SSL+ASR-based in the group (\textrm{II}) of Table~\ref{tab:compare_with_baselines}. The performance of the proposed approach is better than the ASR-based counterpart when transcriptions based on ASR (with a WER of $12.2\%$) are used, but slightly worse than when transcriptions of human annotations are used. Given human annotations are usually not available in a practical ``open response" scenario, we conclude that the proposed SSL+ASR-free approach is superior to both ASR-based and ASR-free baselines in an ``open response" scenario.

\begin{figure}[ht!]
	\centering
	\includegraphics[width=\columnwidth]{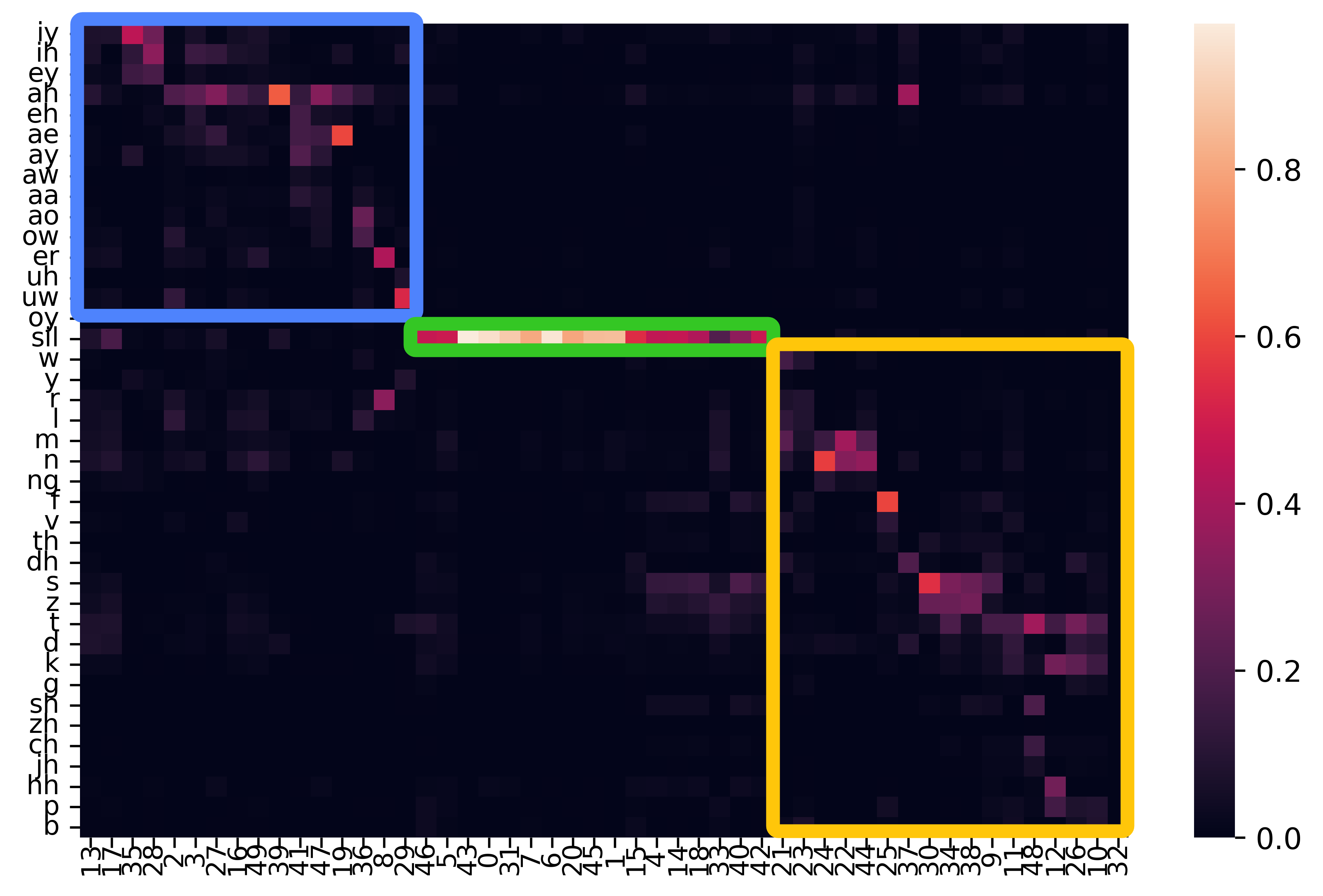}
	\caption{Visualization of the co-occurrence between SSL cluster indexes and phonemes over the training set of ByteQA corpus. Specifically, the conditional probability $P(phoneme|index)$ is plotted. The x-axis shows $50$ different cluster indexes and the y-axis list 40 standard English phonemes, including an additional ``sil" symbol representing the silence (in green row). }
	\label{fig:idx2phone}
	\vspace{-3mm}
\end{figure}

\subsubsection{Analysis of the SSL cluster index}
To evaluate the results of clustering based on SSL features, 
the co-occurrences of cluster indexes and phonemes are counted over the whole training set of ByteQA (about a total of 2.76M indexes) \cite{baevski2020wav2vec}. For each phoneme, the number of time frames that falls into the interval of that phoneme is counted for each cluster index. The phoneme interval is determined by forced alignment with corresponding human transcriptions. As a result, a co-occurrence matrix  $\mathbf{M}_{cooccur}$ of size $40 \times 50$  is obtained, where $50$ is the number of cluster indexes  and $40$ is the sum of $39$ standard English phonemes and an additional silence symbol ``sil".


Figure \ref{fig:idx2phone} gives a visualization plot of conditional probability $P(phoneme|index)$, which is obtained by normalizing $\mathbf{M}_{cooccur}$ with the total occurrence number of each index. The green bounding box represents the result for ``sil", while the blue and yellow boxes represent vowels and consonants, respectively.  It can be seen that most of the cluster indexes appear to specialize in some specific phonemes. The sound ``sil", which accounts for around $19\%$ of speech data and also serves as an important indicator of speech disfluency, corresponds to a wide range of clusters indexes.

The results in general imply that the clusters obtained in the current method can be related to the underlying phonetic units. As such, a sequence of cluster indexes can implicitly represent phoneme boundaries in given non-native speech. This may serve as evidence to explain why the proposed approach can achieve such superior performance without requiring forced alignment to obtain the explicit phoneme boundary.

\subsection{Experimental Results in the ``Read Aloud" scenario}
To verify the performance of the proposed approach in the ``read aloud" scenario, we further conduct experiments on two more datasets, ByteRead and Speechocean762. ByteRead is another internal dataset collected under the "read aloud" scenario and a detailed description is given in~\cite{fu22b_interspeech}. Speechocean762 is an open-sourced speech assessment corpus with 5,000 utterances collected from 250 speakers \cite{zhang2021speechocean762}. For ByteRead, the train/dev/test sets are split as 10k/2k/2k and train/test sets are equally split for Speechocean762. It is ensured that there is no speaker overlap among all the splits for each corpus. The results are given in Table~\ref{tab:read_aloud}.

\begin{table}[h!]
\centering
\caption{The PCC performance in the ``read aloud" scenario.}
\resizebox{\columnwidth}{!}{
\begin{tabular}{c|cc|cc}
\toprule
Methods                                                                             & ASR-free           & Use Text           & ByteRead               & Speechocean762         \\ \midrule
Deep feature+ASR-based\cite{fu22b_interspeech}  & \xmark & \cmark & 0.817 & -     \\
SSL \& Text \cite{kim2022automatic}                                                      & \cmark & \cmark & -     & 0.780 \\ \midrule
SSL + ASR-based                                                    & \xmark & \cmark & 0.847 & 0.804 \\
\textbf{Proposed SSL+ASR-free}  & \cmark & \xmark & \textbf{0.828} & \textbf{0.795}                      \\ \bottomrule
\end{tabular}
}
\label{tab:read_aloud}
\end{table}
As can be seen in Table~\ref{tab:read_aloud}, the system with the setting of ``SSL + ASR-based", i.e., SSL features with forced-alignment, achieves the best performance, followed by the proposed approach and then the other two approaches.
The recent work ``SSL \& Text" \cite{kim2022automatic} developed a scorer by modeling the SSL speech representations and prompt text information as two separate streams. Note that different sizes and types of SSL models were tried and the best-reported performance comes from the non-native ASR fine-tuned HuBERT Large.   
Given the fact that no actual transcriptions of any form are needed/used in the proposed approach, the results demonstrate its effectiveness and potential superiority over ASR-based approaches in automatic speech fluency assessment.


\section{Conclusion}
\label{sec:conclusion}
SSL features extracted from wav2vec 2.0 are shown to possess the ability of phoneme discrimination on non-native datasets in this paper.
Hence, the cluster sequence information originating from SSL features implicitly contains phoneme boundary information,
which is required in modeling fluency/disfluency characteristics.
When taking a sequence of frame-level SSL features and corresponding cluster index embeddings as input to the BLSTM-based scorer, we show that the proposed method significantly outperforms previous ASR-free and ASR-based methods in both ``read aloud" and ``open response" scenarios. Moreover, our method's performance is competitive even to its ASR-based counterpart when using SSL as input. Other types of SSL features, e.g., HuBERT, can be investigated. The efficiency of this  frame-level system will be taken into consideration for further improvement. 

\bibliographystyle{style/IEEEbib}
\small
\bibliography{ref}

\begin{thebibliography}{10}

\bibitem{housen2009complexity}
Alex Housen and Folkert Kuiken,
\newblock ``Complexity, accuracy, and fluency in second language acquisition,''
\newblock {\em Applied linguistics}, vol. 30, no. 4, pp. 461--473, 2009.

\bibitem{kormos2004exploring}
Judit Kormos and Mariann D{\'e}nes,
\newblock ``Exploring measures and perceptions of fluency in the speech of
  second language learners,''
\newblock {\em System}, vol. 32, no. 2, pp. 145--164, 2004.

\bibitem{cucchiarini2002quantitative}
Catia Cucchiarini, Helmer Strik, and Lou Boves,
\newblock ``Quantitative assessment of second language learners’ fluency:
  Comparisons between read and spontaneous speech,''
\newblock {\em the Journal of the Acoustical Society of America}, vol. 111, no.
  6, pp. 2862--2873, 2002.

\bibitem{ellis1994study}
Rod Ellis and Rod~R Ellis,
\newblock {\em The study of second language acquisition},
\newblock Oxford University, 1994.

\bibitem{meng2009developing}
Helen Meng,
\newblock ``Developing speech recognition and synthesis technologies to support
  computer-aided pronunciation training for chinese learners of english,''
\newblock in {\em Proceedings of the 23rd Pacific Asia Conference on Language,
  Information and Computation, Volume 1}, 2009, pp. 40--42.

\bibitem{loukina2019automated}
Anastassia Loukina, Beata~Beigman Klebanov, Patrick~L Lange, Yao Qian, Binod
  Gyawali, Nitin Madnani, Abhinav Misra, Klaus Zechner, Zuowei Wang, and John
  Sabatini,
\newblock ``Automated estimation of oral reading fluency during summer camp
  e-book reading with myturntoread.,''
\newblock in {\em Proc. of Interspeech}, 2019, pp. 21--25.

\bibitem{deng2020automatic}
Huaijin Deng, Youchao Lin, Takehito Utsuro, Akio Kobayashi, Hiromitsu
  Nishizaki, and Junichi Hoshino,
\newblock ``Automatic fluency evaluation of spontaneous speech using
  disfluency-based features,''
\newblock in {\em Proc. of ICASSP}. IEEE, 2020, pp. 9239--9243.

\bibitem{fontan2020using}
Lionel Fontan, Maxime Le~Coz, and Charlotte Alazard,
\newblock ``Using the forward-backward divergence segmentation algorithm and a
  neural network to predict l2 speech fluency,''
\newblock in {\em Proc. 10th International Conference on Speech Prosody}, 2020,
  pp. 925--929.

\bibitem{van2015automatically}
Rogier~C van Dalen, Kate~M Knill, and Mark~JF Gales,
\newblock ``Automatically grading learners' english using a gaussian
  process.,''
\newblock in {\em SLaTE}, 2015, pp. 7--12.

\bibitem{kjellgren2016convolutional}
Fredrik Kjellgren and J~Nordstrom,
\newblock ``Convolutional neural networks for semantic classification of fluent
  speech phone calls,''
\newblock {\em Proc. 6th SLTC}, 2016.

\bibitem{mao2019nn}
Shaoguang Mao, Zhiyong Wu, Jingshuai Jiang, Peiyun Liu, and Frank~K Soong,
\newblock ``Nn-based ordinal regression for assessing fluency of esl speech,''
\newblock in {\em Proc. of ICASSP}. IEEE, 2019, pp. 7420--7424.

\bibitem{zhang2021multilingual}
Huayun Zhang, Ke~Shi, and Nancy~F Chen,
\newblock ``Multilingual speech evaluation: Case studies on english, malay and
  tamil,''
\newblock {\em arXiv preprint arXiv:2107.03675}, 2021.

\bibitem{mahesha2013approach}
P~Mahesha and DS~Vinod,
\newblock ``An approach for classification of dysfluent and fluent speech using
  k-nn and svm,''
\newblock {\em arXiv preprint arXiv:1301.1932}, 2013.

\bibitem{gong2022transformer}
Yuan Gong, Ziyi Chen, Iek-Heng Chu, Peng Chang, and James Glass,
\newblock ``Transformer-based multi-aspect multi-granularity non-native english
  speaker pronunciation assessment,''
\newblock in {\em ICASSP 2022-2022 IEEE International Conference on Acoustics,
  Speech and Signal Processing (ICASSP)}. IEEE, 2022, pp. 7262--7266.

\bibitem{chao20223m}
Fu-An Chao, Tien-Hong Lo, Tzu-I Wu, Yao-Ting Sung, and Berlin Chen,
\newblock ``3m: An effective multi-view, multi-granularity, and multi-aspect
  modeling approach to english pronunciation assessment,''
\newblock {\em arXiv preprint arXiv:2208.09110}, 2022.

\bibitem{fu22b_interspeech}
Kaiqi Fu, Shaojun Gao, Xiaohai Tian, Wei Li, and MA~Zejun,
\newblock ``{Using Fluency Representation Learned from Sequential Raw Features
  for Improving Non-native Fluency Scoring},''
\newblock in {\em Proc. of Interspeech 2022}, pp. 4337--4341.

\bibitem{chung2017spoken}
Hoon Chung, Yun~Kyung Lee, Sung~Joo Lee, and Jeon~Gue Park,
\newblock ``Spoken english fluency scoring using convolutional neural
  networks,''
\newblock in {\em Proc. of O-COCOSDA}, 2017, pp. 1--6.

\bibitem{andre1988new}
Regine Andre-Obrecht,
\newblock ``A new statistical approach for the automatic segmentation of
  continuous speech signals,''
\newblock {\em IEEE Transactions on Acoustics, Speech, and Signal Processing},
  vol. 36, no. 1, pp. 29--40, 1988.

\bibitem{baevski2020wav2vec}
Alexei Baevski, Yuhao Zhou, Abdelrahman Mohamed, and Michael Auli,
\newblock ``wav2vec 2.0: A framework for self-supervised learning of speech
  representations,''
\newblock {\em Advances in Neural Information Processing Systems}, vol. 33, pp.
  12449--12460, 2020.

\bibitem{hsu2021hubert}
Wei-Ning Hsu, Benjamin Bolte, Yao-Hung~Hubert Tsai, Kushal Lakhotia, Ruslan
  Salakhutdinov, and Abdelrahman Mohamed,
\newblock ``Hubert: Self-supervised speech representation learning by masked
  prediction of hidden units,''
\newblock {\em IEEE/ACM Transactions on Audio, Speech, and Language
  Processing}, vol. 29, pp. 3451--3460, 2021.

\bibitem{mohamed2022self}
Abdelrahman Mohamed, Hung-yi Lee, Lasse Borgholt, Jakob~D Havtorn, Joakim Edin,
  Christian Igel, Katrin Kirchhoff, Shang-Wen Li, Karen Livescu, Lars
  Maal{\o}e, et~al.,
\newblock ``Self-supervised speech representation learning: A review,''
\newblock {\em arXiv preprint arXiv:2205.10643}, 2022.

\bibitem{baevski2021unsupervised}
Alexei Baevski, Wei-Ning Hsu, Alexis Conneau, and Michael Auli,
\newblock ``Unsupervised speech recognition,''
\newblock {\em Advances in Neural Information Processing Systems}, vol. 34, pp.
  27826--27839, 2021.

\bibitem{zhang2018deep}
Shiliang Zhang, Ming Lei, Zhijie Yan, and Lirong Dai,
\newblock ``Deep-fsmn for large vocabulary continuous speech recognition,''
\newblock in {\em Proc. of ICASSP}. IEEE, 2018, pp. 5869--5873.

\bibitem{panayotov2015librispeech}
Vassil Panayotov, Guoguo Chen, Daniel Povey, and Sanjeev Khudanpur,
\newblock ``Librispeech: an asr corpus based on public domain audio books,''
\newblock in {\em Proc. of ICASSP}. IEEE, 2015, pp. 5206--5210.

\bibitem{graves2012sequence}
Alex Graves,
\newblock ``Sequence transduction with recurrent neural networks,''
\newblock {\em arXiv preprint arXiv:1211.3711}, 2012.

\bibitem{kingma2014adam}
Diederik~P Kingma and Jimmy Ba,
\newblock ``Adam: A method for stochastic optimization,''
\newblock {\em arXiv preprint arXiv:1412.6980}, 2014.

\bibitem{zhang2021speechocean762}
Junbo Zhang, Zhiwen Zhang, Yongqing Wang, Zhiyong Yan, Qiong Song, Yukai Huang,
  Ke~Li, Daniel Povey, and Yujun Wang,
\newblock ``speechocean762: An open-source non-native english speech corpus for
  pronunciation assessment,''
\newblock {\em arXiv preprint arXiv:2104.01378}, 2021.

\bibitem{kim2022automatic}
Eesung Kim, Jae-Jin Jeon, Hyeji Seo, and Hoon Kim,
\newblock ``Automatic pronunciation assessment using self-supervised speech
  representation learning,''
\newblock {\em arXiv preprint arXiv:2204.03863}, 2022.

\end{thebibliography}

\end{document}